\documentclass[12pt]{article}
\usepackage{epsfig,amssymb,amsmath,psfrag,hyperref}
\usepackage[utf8]{inputenc}
\usepackage[english]{babel}
 
\usepackage{color}

\hypersetup{
  linktocpage,
  colorlinks  = true, 
  urlcolor    = blue, 
  linkcolor   = blue, 
  citecolor   = red 
}

\newcommand{\cN}{{\cal N}}

\newcommand{\cO}{{\cal O}}
\newcommand{\cM}{{\cal M}}
\newcommand{\cA}{{\cal A}}

\def \be  {\begin{equation}}
\def \ee  {\end{equation}}
\def \ba  {\begin{eqnarray}}
\def \ea  {\end{eqnarray}}

\def \cO{\mathcal{O}}

\newcommand \widebar [1] {\overline{#1}}

\def\x{x}
\def\xb{\bar{x}}
\def\dbar#1{\widebar{D}_{#1}}
\def\me{\mathcal{M}}

\def\l{\lambda}
\def\ll{\lambda^{-\frac{3}{2}}}
\def\lll{\lambda^{-\frac{5}{2}}}

\begin{document}
\thispagestyle{empty}

\null\vskip-12pt \hfill  \\
\null\vskip-12pt \hfill   \\

\vskip2.2truecm
\begin{center}
\vskip 0.2truecm {\Large\bf
{\Large String corrections to AdS amplitudes and the double-trace spectrum of $\mathcal{N}=4$ SYM}
}\\
\vskip 1truecm
{\bf J.~M. Drummond, D.~Nandan, H.~Paul,  K.~S. Rigatos \\
}

\vskip 0.4truecm
 
{\it
 School of Physics and Astronomy, University of Southampton,\\
 Highfield, Southampton SO17 1BJ\\
\vskip .2truecm                        }
\end{center}

\vskip 1truecm 
\centerline{\bf Abstract}\normalsize
We consider $\alpha'$ corrections to four-point correlators of half-BPS operators in $\mathcal{N}=4$ super Yang-Mills theory in the supergravity limit. By demanding the correct behaviour in the flat space limit, we find that the leading $(\alpha')^3$ correction to the Mellin amplitude is fixed for arbitrary charges of the external operators. By considering the mixing of double-trace operators we can find the $(\alpha')^3$ corrections to the double-trace spectrum which we give explicitly for $su(4)$-singlet operators. We observe striking patterns in the corrections to the spectra which hint at their common ten-dimensional origin. By extending the observed patterns and imposing them at order $(\alpha')^5$ we are able to reproduce the recently found result for the correction to the Mellin amplitude for $\langle \mathcal{O}_2 \mathcal{O}_2 \mathcal{O}_p \mathcal{O}_p \rangle$ correlators. By applying a similar logic to the $[0,1,0]$ channel of $su(4)$ we are able to deduce new results for the correlators of the form $\langle \mathcal{O}_2 \mathcal{O}_3 \mathcal{O}_{p-1} \mathcal{O}_p \rangle$.

\medskip

\noindent                                                                                                                                   
\newpage
\setcounter{page}{1}\setcounter{footnote}{0}
\tableofcontents

\section{Introduction and summary}\setcounter{equation}{0}
Recently there has been significant progress in understanding the supergravity limit of correlation functions in $\mathcal{N}=4$ super Yang-Mills theory. In this limit, one first expands around large $N$ while keeping the 't Hooft coupling $\lambda=g^2 N$ fixed, followed by a large $\lambda$ expansion. According to the AdS/CFT correspondence such correlators are equivalent to scattering amplitudes of supergravity states in AdS. In particular, a general formula for the tree-level supergravity contribution to four-point AdS amplitudes with arbitrary single-particle half-BPS states was given in Mellin space in \cite{Rastelli:2016nze,Rastelli:2017udc} which generalises many partial results obtained by various means (see e.g. \cite{Dolan:2006ec,Uruchurtu:2008kp,Uruchurtu:2011wh}).

With explicit tree-level results available there has also been progress in understanding the loop corrections to such supergravity amplitudes \cite{Alday:2017xua,Aprile:2017bgs}. To explore such loop corrections it is necessary to address a mixing problem involving many double-trace operators with the same classical quantum numbers. By analysing sufficient correlators it is possible to obtain the order $1/N^2$ anomalous dimensions for all double-trace operators of arbitrary $su(4)$ quantum numbers \cite{Aprile:2018efk} as well as explicit results for the leading order three-point functions of two half-BPS operators and one double-trace operator \cite{unmixing}.

One may also ask about $\alpha' \sim \lambda^{-\frac{1}{2}}$ corrections to the supergravity results. Such corrections have been addressed in several papers, e.g. \cite{Goncalves:2014ffa,Alday:2018pdi,Binder:2019jwn}. Mellin space is particularly convenient for the analysis of such correlators. Here we will investigate the $\lambda^{-\frac{1}{2}}$ corrections to tree-level supergravity further. We argue that the consistency with the ten-dimensional supersymmetric Virasoro-Shapiro amplitude in the flat space limit completely determines the leading $\lambda^{-\frac{3}{2}}$ correction to all tree-level Mellin amplitudes. With this information available, we will derive the form of the corrections to the anomalous dimensions and three-point functions for double-trace operators in the singlet channel. We find that the corrections to the three-point functions vanish while the anomalous dimensions exhibit a very simple structure. The degeneracy among the $su(4)$ singlet channel operators of a given twist is fully lifted by the order $1/N^2$ tree-level supergravity anomalous dimensions. Of these operators only the lightest at any given twist receives a $\lambda^{-\frac{3}{2}}$ correction. Similar behaviour is exhibited in the $[0,1,0]$ channel. We interpret this feature having  its origins in the ten-dimensional symmetry described in \cite{Caron-Huot:2018kta}. Specifically, we trace the origin of this property to the fact that the $(\alpha')^3 \mathcal{R}^4$ correction in the type IIB superstring effective action can only involve exchanged operators of ten-dimensional spin zero. The fact that features of the ten-dimensional symmetry remain even after including $\alpha'$ corrections is very surprising, as one might think this symmetry is inevitably broken by string corrections.

To go beyond order $\ll$, we make the working assumption that the three-point functions remain uncorrected at the next order $\lambda^{-\frac{5}{2}}$. Then the recent results of \cite{Alday:2018pdi,Binder:2019jwn} allow us to obtain the spectrum of anomalous dimensions in a similar way. The pattern of only the lightest states receiving a correction continues, this time corresponding to the operators of ten-dimensional spins zero and two. This is again consistent with our ten-dimensional interpretation, since the relevant term in the effective action is $(\alpha')^5 \partial^4 \mathcal{R}^4$. It follows that if we take the structure of the spectrum as an assumption together with the vanishing of the corrections to the three-point functions, we can derive constraints on an ansatz for the order $\lambda^{-\frac{5}{2}}$ Mellin amplitudes, in a similar spirit to \cite{Alday:2018pdi,Binder:2019jwn}. We show that proceeding in this way we can fix the $\lambda^{-\frac{5}{2}}$ Mellin amplitudes for the $\langle \mathcal{O}_2 \mathcal{O}_2 \mathcal{O}_p \mathcal{O}_p \rangle$ family of correlators up to a single free parameter consistently with \cite{Alday:2018pdi,Binder:2019jwn}.
We then employ the above techniques and assumptions to derive new constraints on the $\langle \mathcal{O}_2 \mathcal{O}_3 \mathcal{O}_{p-1} \mathcal{O}_p \rangle$ family of correlators, again fixing the result up to a single free parameter.

\section{Half-BPS correlators in the supergravity limit}\label{sec:setup}\setcounter{equation}{0}
We recall that the superconformal half-BPS operators corresponding to single-particle states take the form
\begin{align}
	\cO_p=y^{R_1}\cdots y^{R_p}\text{Tr}\left(\Phi_{R_1}\cdots\Phi_{R_p}\right)+ \ldots
	\label{eq:single_trace_op}
\end{align}
where $y^R$ is a null $so(6)$ vector which picks out the traceless symmetric part, i.e. the $[0,p,0]$ representation of $su(4)$. The dots stand for $1/N^2$ suppressed multi-trace terms which are determined by demanding that $\mathcal{O}_p$ is orthogonal to any multi-trace half-BPS operator $\langle \mathcal{O}_p [\mathcal{O}_{q_1} \ldots \mathcal{O}_{q_n}]\rangle = 0$ \cite{Aprile:2018efk}. The operator $\mathcal{O}_2$ is dual to the graviton supermultiplet and the higher $\mathcal{O}_p$ are dual to the Kaluza-Klein modes of the ten-dimensional graviton compactified on the $S^5$ factor of the AdS${}_5 \times S^5$ background.

Here we are interested in four-point functions of such half-BPS operators,
\be
\langle p_1p_2p_3p_4 \rangle:=\langle \cO_{p_1}\cO_{p_2}\cO_{p_3}\cO_{p_4} \rangle\,.
\ee
Superconformal symmetry places strong constraints on the form of such correlators \cite{Eden:2000bk,Nirschl:2004pa}. If we write the correlator as a sum of its free field theory and interacting contributions, then the interacting contribution takes a particular factorised form,
\be
\langle p_1p_2p_3p_4 \rangle = \langle p_1p_2p_3p_4 \rangle_{\rm free} + \mathcal{P} \, \mathcal{I} \, \mathcal{H}\,.
\ee
The factor $\mathcal{P}$ is given by
\be
\label{Pfactor}
\mathcal{P} = g_{12}^{\frac{p_1+p_2-p_{43}}{2}} g_{14}^{\frac{p_{43}-p_{21}}{2}} g_{24}^{\frac{p_{43}+p_{21}}{2}} g_{34}^{\phantom{\frac{}{}}p_3}\,,
\ee
where $p_{ij}=p_i-p_j$ and the $g_{ij} = y_{ij}^2/x_{ij}^2$ (with $y_{ij}^2 = y_i \cdot y_j$) are propagator factors which carry the conformal weights and $y_i$ the scaling weights of the correlator. The remaining factors $\mathcal{I}$ and $\mathcal{H}$ are then functions of the conformal and $su(4)$ cross-ratios,
\begin{align}
	u&= x \bar x = \frac{x_{12}^2x_{34}^2}{x_{13}^2 x_{24}^2}\,,  &&v=(1-x)(1-\bar x) =  \frac{x_{14}^2x_{23}^2}{x_{13}^2 x_{24}^2}\,, \notag \\ \frac1\sigma&= y \bar y = \frac{y_{12}^2 y_{34}^2}{y_{13}^2 y_{24}^2}\,, && \frac\tau \sigma =(1-y)(1-\bar y) =  \frac{y_{14}^2 y_{23}^2}{y_{13}^2 y_{24}^2}\,. 
	\label{crossratios}
\end{align}
The factor $\mathcal{I}$ is given by
\be
\mathcal{I} = \frac{(x-y)(x-\bar{y})(\bar{x}-y)(\bar{x}-\bar{y})}{(y \bar{y})^2}\,.
\ee
The presence of the four zeros in the numerator is a consequence of the superconformal Ward identities and is a feature of the contributions of unprotected operators in the conformal partial wave expansion of the correlation functions. Only unprotected operators may contribute to the interacting term since these are the only operators whose conformal data depend on the gauge coupling.

The factor $\mathcal{H} = \mathcal{H}(u,v;\sigma,\tau)$ is the only piece which depends on the gauge coupling and hence is the piece which contains the dynamics of the theory. In the supergravity limit of large $N$ and large 't Hooft coupling it admits a double expansion of the form
\begin{align}
	\mathcal{H}=\mathcal{H}^{(0,0)}+a\left(\mathcal{H}^{(1,0)}+\ll\mathcal{H}^{(1,3)}+\lll\mathcal{H}^{(1,5)}+\ldots\right)+O(a^2),
	\label{eq:H_expansion}
\end{align}
where to match previous conventions we use $a= 1/(N^2-1)$ as the expansion parameter instead of $1/N^2$ although here the distinction is not really relevant. The leading term $\mathcal{H}^{(0,0)}$ corresponds to the leading large $N$ disconnected contribution which is only non-zero for correlators of the form $\langle ppqq \rangle$ or those related by crossing symmetry. The terms at order $a$ are the tree-level contributions with $\mathcal{H}^{(1,0)}$ being the tree-level supergravity contribution and the other terms due to string corrections.

As described in \cite{Alday:2017xua,Aprile:2017bgs}, if we perform the combined $(\mathcal{O}_{p_1} \times \mathcal{O}_{p_2})$ and $(\mathcal{O}_{p_3} \times \mathcal{O}_{p_4})$ operator product expansion, the contributions to $\mathcal{H}$ in the supergravity limit are controlled by unprotected double-trace operators of the form $[\mathcal{O}_p \partial_{\ell} \Box^{\tfrac{1}{2}(\tau - p - q)} \mathcal{O}_q]_{[a,b,a]}$. They have classical twist $\tau$, spin $\ell$ and $su(4)$ representation $[a,b,a]$. There are typically many such operators with different values of $p$ and $q$ with same quantum numbers, leading to a mixing problem. Such operators, being unprotected, acquire anomalous dimensions in the double large $N$, large $\lambda$ expansion,
\begin{align}
	\Delta = \Delta^{(0)} +2a\left(\eta^{(0)}+\ll \eta^{(3)}+\lll \eta^{(5)}+\ldots\right)+O(a^2),
	\label{eq:dim_expansion}
\end{align}
where $\Delta^{(0)} =\tau+\ell$ and $\eta^{(n)}$ depends on $\tau$, $\ell$, $a$, $b$ and the degeneracy labels. The superscript in $\eta^{(n)}$ denotes the order in $\l^{-\frac{n}{2}}$. Note that for later convenience $\eta^{(n)}$ is in fact \textit{half} the anomalous dimension, however, for the rest of this paper, we will simply refer to it as the anomalous dimension.

The knowledge of the tree-level supergravity contributions $\mathcal{H}^{(1,0)}$ due to \cite{Rastelli:2016nze,Rastelli:2017udc} allows the mixing between these operators to be resolved leading to a compact formula for their leading anomalous dimensions $\eta^{(0)}$ \cite{Aprile:2018efk}. In fact, for general $su(4)$ channels, not all of the degeneracy is lifted by the first correction to the dimensions. This feature, as well as the surprisingly simple form of the anomalous dimensions, is related to a novel ten-dimensional symmetry which is exhibited in the supergravity tree-level correlators \cite{Caron-Huot:2018kta}. Here we will argue that the ten-dimensional connection is also responsible for striking patterns observed in the structure of the higher corrections $\eta^{(3)}$, $\eta^{(5)}$ etc.

Before describing the spectrum we show that the flat space limit dictates the form of the first string correction $\mathcal{H}^{(1,3)}$.

\section{The general half-BPS amplitude at order $\ll$}\label{sec:lambda_3/2_general_result}\setcounter{equation}{0}
In~\cite{Alday:2018pdi}, it has been demonstrated how matching the Virasoro-Shapiro amplitude by applying the flat space limit to the AdS$_5\times S^5$ Mellin amplitude fully determines the coefficient of the $\langle 22pp \rangle$-family of correlators up to order $\ll$. In fact, at any order in $1/\l$ the coefficient of the (polynomial) Mellin amplitude with leading $s,t\rightarrow\infty$ asymptotics is fixed by matching the corresponding term in the Virasoro-Shapiro amplitude.

In this section, we restrict our attention to the first order in $1/\l$ and explain how the flat space limit together with the formula for the supergravity correlator~\cite{Rastelli:2016nze,Rastelli:2017udc} and its normalisation as derived in~\cite{Aprile:2018efk} can be used to generalise the results for the special cases mentioned above, arriving at a formula for the general half-BPS four-point amplitude $\langle p_1p_2p_3p_4 \rangle$ at $\ll$ for arbitrary external charges. 
\subsection{The Mellin space ansatz}
Tree-level Witten diagrams are most conveniently represented in Mellin space, where the Mellin amplitudes are rational, as it is the case for tree level supergravity (with a prescribed set of poles and residues, corresponding to the exchanged single-trace operators in a certain Witten diagram), or polynomial in case of higher derivative corrections to the interaction vertices. We will thus use the Mellin space formalism for holographic four-point correlators to describe the interacting part $\mathcal{H}_{\{p_i\}}(u,v;\sigma,\tau)$ of a general four-point correlator, where ${\{p_i\}}$ denotes the dependence on the four external charges $(p_1,p_2,p_3,p_4)$.

We use $p_{43}\geq p_{21}\geq 0$ as our convention for the arrangement of charges, and other cases may be obtained by applying crossing transformations. The inverse Mellin transform of the interacting part is then given by\footnote{\label{footnote:reducedMellinamp}What we call $\me$ here is in fact the reduced Mellin amplitude (usually denoted by $\widetilde{\me}$), which is related to the full Mellin amplitude $M$ by
\[M(s,t;\sigma,\tau)=\widehat{R}(u,v;\sigma,\tau)\circ\widetilde{\mathcal{M}}(s,t;\sigma,\tau),\]
where $\widehat{R}$ is a difference operator mimicking the action of the factor $\mathcal{I}$ on the interacting part $\mathcal{H}$. See~\cite{Rastelli:2017udc} for further details, where also a precise definition of the integration contour is given, such that rational parts of the position space result are correctly recovered from the Mellin integrals.}
\begin{align}
	\mathcal{H}_{\{p_i\}}(u,v;\sigma,\tau) = \int_{-i\infty}^{i\infty} \frac{ds}{2}\frac{dt}{2} ~u^{\frac{s}{2}-\frac{p_{43}}{2}}v^{\frac{t}{2}-\frac{p_2+p_3}{2}}\me_{\{p_i\}}(s,t;\sigma,\tau)\Gamma_{\{p_i\}}(s,t),
\label{eq:mellintrafo}
\end{align}
where the string of six $\Gamma$-functions is defined as
\begin{align}
	\Gamma_{\{p_i\}}(s,t)=\prod_{i<j}\Gamma\left[c_{ij}\right],
\label{eq:RastelliGammas}
\end{align}
with the Mellin space parametrisation $c_{ij}=c_{ji}$ given by\footnote{We should warn the reader that the variable $u$ is being used to denote two different quantities here: it is the ordinary conformal cross-ratio as defined in equation~\eqref{crossratios}, but in the context of Mellin amplitudes we also use it as one of the usual Mellin variables $(s,t,u)$ obeying (\ref{eq:Mandelstam_constraint_in_Mellin_space}). We hope the context will make the distinction clear.}
\begin{align}
\begin{split}
	c_{12}&=-\frac{s}{2}+\frac{p_1+p_2}{2},\qquad
	c_{13}=-\frac{u}{2}+\frac{p_1+p_3}{2},\qquad
	c_{14}=-\frac{t}{2}+\frac{p_1+p_4}{2},\\ 	c_{23}&=-\frac{t}{2}+\frac{p_2+p_3}{2},\qquad
	c_{24}=-\frac{u}{2}+\frac{p_2+p_4}{2},\qquad
	c_{34}=-\frac{s}{2}+\frac{p_3+p_4}{2}.
\end{split}
\label{eq:MellinParam}
\end{align}
Note that the Mellin space Mandelstam variables $(s,t,u)$ satisfy the constraint
\begin{align}
	s+t+u=p_1+p_2+p_3+p_4-4.
	\label{eq:Mandelstam_constraint_in_Mellin_space}
\end{align}
Analogous to the double expansion of the interacting part $\mathcal{H}_{\{p_i\}}$ in equation~\eqref{eq:H_expansion}, we expand the corresponding order $a$ Mellin amplitude in $1/\l$ according to
\begin{align}
	\me_{\{p_i\}} = \me_{\{p_i\}}^{(1,0)} + \ll \me_{\{p_i\}}^{(1,3)} + \lll \me_{\{p_i\}}^{(1,5)} + \ldots.
	\label{eq:lambda_expansion_mellin}
\end{align}
The supergravity Mellin amplitude $\me^{(1,0)}_{\{p_i\}}(s,t;\sigma,\tau)$ for arbitrary external charges has been conjectured in~\cite{Rastelli:2016nze,Rastelli:2017udc} up to an undetermined overall normalisation $\mathcal{N}_{p_1p_2p_3p_4}$, their result being
\begin{align}
	\me_{\{p_i\}}^{(1,0)} = \mathcal{N}_{p_1p_2p_3p_4}\sum_{i,j\geq0}\frac{a_{ijk}\sigma^i\tau^j}{(s-\tilde{s}+2k)(t-\tilde{t}+2j)(u-\tilde{u}+2i)},
	\label{eq:rastelli_formula}
\end{align}
where $k=p_3+\min\{0,\frac{p_1+p_2-p_3-p_4}{2}\}-i-j-2$ and the range of $i,j$ is such that $k\geq0$ in the sum. Furthermore, we define
\begin{align}
\begin{split}
	\tilde{s}&=\min\left\{p_1+p_2,p_3+p_4\right\}-2,\\
	\tilde{t}&=p_2+p_3-2,\\
	\tilde{u}&=p_1+p_3-2.\\
\end{split}	
\end{align}
The overall normalisation $\mathcal{N}_{p_1p_2p_3p_4}$ was subsequently determined in~\cite{Aprile:2018efk}, and it combines nicely with the factor $a_{ijk}$ into $\mathcal{N}_{ijk}\equiv\mathcal{N}_{p_1p_2p_3p_4} a_{ijk}$, given by
\begin{align}
	\mathcal{N}_{ijk} = \frac{1}{i!j!k!} \frac{8p_1p_2p_3p_4}{\big(\frac{p_{43}+p_{21}}{2}+i\big)!\big(\frac{p_{43}-p_{21}}{2}+j\big)!\big(\frac{|p_1+p_2-p_3-p_4|}{2}+k\big)!}.
	\label{eq:sugra_normalisation}
\end{align}
For future convenience, we define $B^{sugra}_{\{p_i\}}(\sigma,\tau)$ as the coefficient of the supergravity Mellin amplitude in the large $s,t$ limit by
\begin{align}
	B^{sugra}_{\{p_i\}}(\sigma,\tau) = \sum_{i,j\geq0}\mathcal{N}_{ijk}\sigma^i\tau^j.
	\label{eq:sugra_coefficient}
\end{align}
Let us now turn our attention to adding string corrections to the supergravity result~(\ref{eq:rastelli_formula}), as already indicated in the expansion~\eqref{eq:lambda_expansion_mellin}. These $1/\l$ corrections descend from higher derivative interaction terms in the AdS$_5\times S^5$ effective action, the first two being $\mathcal{R}^4$ at order $\ll$ and $\partial^4\mathcal{R}^4$ at order $\lll$, respectively. In Mellin space, the analytic structure of tree-level Witten diagrams dictates that for a general correction term of the schematic form $\partial^{2n}\mathcal{R}^4$, the corresponding Mellin amplitude is simply a polynomial of degree $n$, together with all subleading polynomial amplitudes coming from terms in 10d with legs on $S^5$~\cite{Penedones:2010ue,Fitzpatrick:2011hu,Alday:2014tsa,Goncalves:2014ffa,Alday:2018pdi}.\footnote{The tree-level corrections to the supergravity Mellin amplitude are polynomial since they correspond to corrections due to unprotected double-trace operators, whose poles are already present in the gamma functions in (\ref{eq:RastelliGammas}). The bound on the polynomial comes from considering the flat space limit and moreover the coefficients of the leading Mellin amplitudes can be fixed by comparing against the flat space 10d IIB closed superstring amplitude, as we will discuss in the next section.} In the same spirit as \cite{Alday:2018pdi,Binder:2019jwn} we can thus make an ansatz of the form
\begin{align}
\begin{split}
	\me_{\{p_i\}}^{(1,3)} &= B^4_4\me_{\{p_i\}}^4,\\
	\me_{\{p_i\}}^{(1,5)} &= B^6_{6,j}\me_{\{p_i\}}^{6,j}+B^6_{5,j}\me_{\{p_i\}}^{5,j}+B^6_4\me_{\{p_i\}}^4,
\end{split}\label{eq:mellin_expansions}
\end{align}
where $\me_{\{p_i\}}^n$ are polynomial Mellin amplitudes of degree $(n-4)$ in the Mellin variables $(s,t,u)$ and their coefficients $B^k_n\equiv B^k_n(\sigma,\tau;\{p_i\})$ are in general functions of the internal cross-ratios $(\sigma,\tau)$ as well as the four external charges.\footnote{The extra index $j$ in $\me_{\{p_i\}}^{n,j}$ and $B^k_{n,j}$ is used when there exists more than one independent Mellin polynomial of degree $(n-4)$.} Explicit expressions for the basis of Mellin polynomials for the $\langle 22pp \rangle$ and $\langle 23(p-1)p \rangle$ families of correlators are given in section~\ref{sec:unmixing_lambda_5/2}. Note that adding a term of the same form as the supergravity amplitude $\me_{\{p_i\}}^1$ in the above ansatz~\eqref{eq:mellin_expansions} is precluded since it is not polynomial. Including such a term would spoil the cancellation of excited string states at low twist between the free theory contribution and $\mathcal{H}$ \cite{Dolan:2001tt}. 

The first polynomial correction, associated with the $\mathcal{R}^4$ vertex, is given by the constant Mellin amplitude
\begin{align}
	\me_{\{p_i\}}^4 = 1,
	\label{eq:mellin_amplitude_lambda_3/2}
\end{align}
which is trivially crossing symmetric on its own. At order $\ll$, where only $\me_{\{p_i\}}^4$ contributes, it will turn out that the simplicity of $\me_{\{p_i\}}^4$ together with the knowledge of the supergravity result~(\ref{eq:rastelli_formula}) and its normalisation~(\ref{eq:sugra_normalisation}) is enough to fix its coefficient $B^4_4$ for all external charges.
\subsection{The flat space limit}
Let us briefly explain the method of matching the flat space limit, which was first motivated by Penedones~\cite{Penedones:2010ue} and explored further in~\cite{Fitzpatrick:2011hu}. This method was first applied to the $\langle 2222 \rangle$ correlator by Gon\c{c}alves~\cite{Goncalves:2014ffa}, and more recently extended to the $\langle 22pp \rangle$ family of correlators~\cite{Alday:2018pdi,Binder:2019jwn}. Their discussion is based on previous work in AdS$_7\times S^4$~\cite{Chester:2018aca,Chester:2018dga}, whose logic we will follow here to extend the previous results to the general correlator $\langle p_1p_2p_3p_4 \rangle$ with non-trivial $(\sigma,\tau)$ dependence.

Penedones defines the following relation between the flat space and Mellin amplitudes:
\begin{equation}
M(s_{ij})= \frac{R^{n(1-d)/2+d+1}}{\Gamma(\Sigma-d/2)}\int_0^\infty d\beta \beta^{\Sigma-d/2-1}e^{-\beta}\cA_{\text{Flat}}\big(\frac{2\beta}{R^2} s_{ij}\big), \quad s_{ij} \gg 1,
\label{eq:Penedonesformula}
\end{equation}
where $ \Sigma= \frac{p_1+p_2+p_3+p_4}{2} $ is half the sum of the dimensions of the external operators for the four-point function under consideration with $ n=4 $, $ d=4 $ and $ R $ being the radius of the AdS space, such that  $\alpha' = l_s^2 =\lambda^{-\frac{1}{2}}R^2 $. We want to start from  this $ 10 $ dimensional expression and restrict the  kinematics to the $ 5 $ dimensional AdS subspace, or rather $ \mathbb{R} _5 \equiv AdS_5\vert_{R\rightarrow \infty} $, by integrating over the $ 5 $ dimensional wavefunction dual to the internal Kaluza-Klein modes~\cite{Chester:2018dga}. Equation \eqref{eq:Penedonesformula} can now be inverted as,
\begin{equation}
\cA_{\text{Flat}}(s,t)= \lim_{R \rightarrow \infty} \Gamma(\Sigma-d/2)R^6\Bigl( \int_{S^5}d^5x\sqrt{g}\prod_{i=1}^n\Psi_{\eta_i}^{\cO_i}(\vec{n})\Bigr)\int _{-i\infty}^{+i \infty} \frac{d \alpha}{2 \pi i} \alpha^{-(\Sigma-d/2)} e^\alpha M\Bigl(\frac{R^2}{2\alpha} s,\frac{R^2}{2\alpha} t\Bigr).
\label{eq:FlatspaceMainformula}
\end{equation}
By taking the flat space limit, we should match against the type IIB closed string theory scattering amplitude of four super-gravitons, which admits an expansion in the string coupling $g_s$ (see e.g.~\cite{Polchinski:1998rr})
\begin{align}
\mathcal{A}_{\text{Flat}}&=\mathcal{A}_{\text{sugra}}f(s,t),\quad \text{with}\nonumber\\ f(s,t)&=-stu~\frac{\lambda^{-\frac{3}{2}}R^6}{64 }~\frac{\Gamma\big(-\frac{s}{4}\lambda^{-\frac{1}{2}}R^2\big)\Gamma\big(-\frac{t}{4}\lambda^{-\frac{1}{2}}R^2\big)\Gamma\big(-\frac{u}{4}\lambda^{-\frac{1}{2}}R^2\big)}{\Gamma\big(1+\frac{s}{4}\lambda^{-\frac{1}{2}}R^2\big)\Gamma\big(1+\frac{t}{4}\lambda^{-\frac{1}{2}}R^2\big)\Gamma\big(1+\frac{u}{4}\lambda^{-\frac{1}{2}}R^2\big)}+O(g_s^2),
\label{eq:virasoroshapiro}
\end{align}
where $\mathcal{A}_{\text{sugra}}$ is the tree-level supergravity scattering amplitude, $(s,t,u)$ are the usual 10d Mandelstam invariants obeying $s+t+u=0$. 
For our purposes it is enough to consider the leading term in $g_s$, as it corresponds to the leading large $N$ result in $\mathcal{N}=4$ SYM in the 't Hooft limit with fixed $\l$. Worldsheets with genus one and higher contribute to higher orders in $g_s$, corresponding to subleading $1/N$ corrections in the CFT (see~\cite{Alday:2018pdi,Alday:2018kkw} for applications to string corrections to the correlator $\langle 2222 \rangle$ at order $1/N^4$).
A further expansion of $f(s,t)$ in $1/\l$ gives
\begin{align}
f(s,t)=\left(1+stu\frac{\zeta_3}{32}\cdot\lambda^{-\frac{3}{2}}R^6+stu(s^2+t^2+u^2)\frac{\zeta_5}{1024}\cdot\lambda^{-\frac{5}{2}}R^{10}+\ldots\right) +O(g_s^2).
\label{eq:virasoroexpansion}
\end{align}
As a result  of the superconformal Ward Identities the Mellin amplitude $M$ in \eqref{eq:FlatspaceMainformula} is related to the reduced Mellin amplitude $\me$ defined in equation \eqref{eq:mellintrafo}, as mentioned in footnote~\eqref{footnote:reducedMellinamp}. In the flat space limit this is given by
\begin{align}
M(s,t)\rightarrow \frac{1}{16}\big(t^2u^2+s^2t^2\sigma^2+s^2u^2\tau^2+2s^2tu\sigma\tau+2st^2u\sigma+2stu^2\tau\big)\cM(s,t).
\label{eq:MandcurlyMrelation}
\end{align}
Using the above definition in \eqref{eq:FlatspaceMainformula} and comparing with \eqref{eq:virasoroshapiro} we have
\begin{equation}
f(s,t)=  \frac{\Gamma(\Sigma-\frac{d}{2})}{16 \cN_\cA}\lim_{R \rightarrow \infty}R^{14}\int _{-i\infty}^{+i \infty} \frac{d \alpha}{2 \pi i} \alpha^{-(\Sigma-\frac{d}{2}+4)} e^\alpha \cM\Bigl(\frac{R^2}{2\alpha}s,\frac{R^2}{2\alpha}t\Bigr),
\label{eq:flatspaceCompare}
\end{equation}
where $ \cN_\cA $ is the normalisation constant that takes into account the flat-space supergravity amplitude, such that the first term in the Mellin expansion matches it. Note that $\cN_\cA$ has a non-trivial dependence on $(\sigma,\tau)$, linked to the supergravity coefficient $B^{sugra}_{\{p_i\}}(\sigma,\tau)$, see equation~\eqref{eq:sugra_coefficient}. Using the ansatz for the expansion of $ \cM(s,t) $ in $a=\frac{1}{N^2-1} $ and the unfixed coefficients $ B_n^k $, at $ \cO(a) $ we get
\begin{eqnarray}
\cN_\cA&=&  \frac{2\Gamma(\Sigma-\frac{d}{2})}{(stu)}\lim_{R \rightarrow \infty} a R^{8}	B^{sugra}_{\{p_i\}}(\sigma,\tau)\int _{-i\infty}^{+i \infty} \frac{d \alpha}{2 \pi i} \alpha^{-(\Sigma-\frac{d}{2}+1)} e^\alpha \nonumber \\
&= & \frac{128 \pi^2 g_s^2 l_s^8\Gamma(\Sigma-\frac{d}{2})}{ (stu)}\frac{	B^{sugra}_{\{p_i\}}(\sigma,\tau)}{\Gamma(\Sigma-\frac{d}{2}+1)}.
\label{eq:flatspacenormalization}
\end{eqnarray}
Now, from the full expansion of the rhs in equation \eqref{eq:flatspaceCompare} we find
\begin{eqnarray}
f(s,t)&=&\frac{(stu)}{ 	B^{sugra}_{\{p_i\}}(\sigma,\tau)}\Biggl[ \frac{	B^{sugra}_{\{p_i\}}(\sigma,\tau)}{(stu)} +\lambda^{-\frac{3}{2}}R^6\frac{B_4^4(\sigma,\tau) }{2^3 (\Sigma-\frac{d}{2}+1)_3} +\ldots \Biggr].
\label{eq:fullexpansionFst}
\end{eqnarray}
Comparing \eqref{eq:fullexpansionFst} with the expansion of the string amplitude in \eqref{eq:virasoroexpansion} we can determine the unfixed coefficients of the leading polynomial amplitudes, in particular the coefficient $B^4_4(\sigma,\tau)$ of the Mellin amplitude at order $\ll$ is given by
\begin{equation}
B_4^4(\sigma,\tau)= \frac{(\Sigma-\frac{d}{2}+1)_3\zeta_3}{4}~	B^{sugra}_{\{p_i\}}(\sigma,\tau).
\label{eq:b44_coefficient}
\end{equation}
Note that this result fully determines the coefficient of the $\ll$ Mellin amplitude $\me_{\{p_i\}}^4$ as a function of the charges $(p_1,p_2,p_3,p_4)$. This result relies on the fact that the non-trivial dependence on the internal cross-ratios is fully captured by the supergravity coefficient $B^{sugra}_{\{p_i\}}(\sigma,\tau)$, with the remaining part of the coefficient depending only on the sum of the charges $\Sigma$. This is a consequence of $(\sigma,\tau)$ not being affected by the flat space limit, since we restrict the 10 dimensional momenta to a 5 dimensional subspace while the remaining 5 dimensions (in this case compactified to $S^5$ with $(\sigma,\tau)$ being the coordinates of the spherical harmonics) do not participate in a tree-level scattering process.

\subsection{Result for the Mellin amplitude}
Let us now present the formula for the four-point correlator $\langle p_1p_2p_3p_4 \rangle$ at $\ll$ for arbitrary external charges, the main result of this section. The derivation relies on three ingredients: the simplicity of the Mellin amplitude~(\ref{eq:mellin_expansions}) at order $\ll$, the knowledge of the supergravity result, its normalisation in particular, and the adaptation of the flat space limit to correlators with general external charges.

Without further delay, using the relation~\eqref{eq:b44_coefficient} obtained from the flat space limit, we are led to the compact result
\begin{align}
	\me_{\{p_i\}}^{(1,3)} = \frac{(\Sigma-1)_3\zeta_3}{4}~B^{sugra}_{\{p_i\}}(\sigma,\tau),
\end{align}
from which we easily obtain the explicit position space expression by performing the inverse Mellin transform, resulting in
\begin{align}
	\mathcal{H}_{\{p_i\}}^{(1,3)}= \frac{(\Sigma-1)_3\zeta_3}{4}~B^{sugra}_{\{p_i\}}(\sigma,\tau)~ u^{\frac{p_1+p_2+p_3-p_4}{2}} \dbar{p_1+2,p_2+2,p_3+2,p_4+2}(u,v).
		\label{eq:lambda32generalformula}
\end{align}
For convenience, let us repeat the definition
\begin{align}
	B^{sugra}_{\{p_i\}}(\sigma,\tau) = \sum_{i,j\geq0}\mathcal{N}_{ijk}\sigma^i\tau^j,
\end{align}
where $\mathcal{N}_{ijk}$ was introduced in equation~\eqref{eq:sugra_normalisation}.

Our formula is consistent with the results for $\langle 2222 \rangle$~\cite{Goncalves:2014ffa} and $\langle 22pp \rangle$~\cite{Alday:2018pdi} and by construction obeys the correct crossing transformation properties. We checked explicitly for many cases that, upon decomposing into conformal blocks, our result~(\ref{eq:lambda32generalformula}) contributes to spin 0 only, as expected from the $\mathcal{R}^4$ correction term.

In the next section, we will use this result to initiate the study of anomalous dimensions at order $\ll$ for the singlet and the $[0,1,0]$ channel.
\section{Unmixing the $\ll$ double-trace spectrum}\label{sec:unmixing_lambda_3/2}\setcounter{equation}{0}
As mentioned in section~\ref{sec:setup}, the spectrum of exchanged operators in the OPE at leading order in $1/N$ consists of a set of degenerate double-trace operators. In this section, we describe how to resolve the mixing of these operators at order $\ll$, obtaining analytical formulae for their anomalous dimensions $\eta^{(3)}$.

We follow the approach developed in~\cite{unmixing}, where an OPE analysis was used to determine the spectrum of supergravity anomalous dimensions, denoted by $\eta^{(0)}$ in equation~(\ref{eq:dim_expansion}). First, we explain how the same method can be used to unmix further string corrections to the spectrum. Then we apply the procedure to the singlet and $[0,1,0]$ channel of the $su(4)$ R-symmetry group and compute the anomalous dimensions $\eta^{(3)}|_{[000]}$ and $\eta^{(3)}|_{[010]}$, revealing a surprisingly simple structure. Lastly, we provide an intuitive 10 dimensional explanation of the observed pattern of anomalous dimensions, which can be used to make further predictions about the spectrum induced by higher derivative string corrections. 
\subsection{The unmixing equations}\label{sec:unmixing_equations}
For simplicity, we restrict the discussion given here to the singlet channel, for which the relevant set of correlators is the $\langle ppqq \rangle$-family.\footnote{\label{footnote:ourpapers}For a detailed explanation and a generalisation to all $su(4)$ channels of the form $[a,b,a]$ we refer the reader to the articles~\cite{Aprile:2017qoy,Aprile:2018efk}.} In this channel, for every half-twist $t=\tau/2$ and spin $\ell$ there are $t-1$ degenerate operators which we label by $i=1,\ldots,t-1$:
\begin{align}
	\big\{ K_i \big\} = \big\{ \cO_2\square^{t-2}\partial^\ell\cO_2,\cO_3\square^{t-3}\partial^\ell\cO_3,\ldots,\cO_t\square^0\partial^\ell\cO_t \big\}.
\end{align}
The interacting part of the correlator, $\mathcal{H}_{ppqq}$, admits an OPE decomposition into long superconformal blocks (see \cite{Nirschl:2004pa,Dolan:2001tt,Arutyunov:2002fh,Dolan:2004iy,Doobary:2015gia} and references therein). Projecting onto the singlet channel, we have the decomposition
\begin{align}
	\mathcal{H}_{ppqq}(u,v)|_{[0,0,0]} = \sum_{t,\ell} A_{t,\ell} G_{t,\ell}(u,v),
	\label{eq:block_deco}
\end{align}
where $G_{t,\ell}(u,v)$ is given by the usual four-dimensional conformal block with a shift of four in its dimension~\cite{Dolan:2000ut,Dolan:2003hv}:
\begin{align}\label{eq:conformal_block}
	G_{t,\ell}(u,v) = (-1)^\ell u^t~\frac{\x^{\ell+1}F_{t+\ell+2}(\x)F_{t+1}(\xb)-\xb^{\ell+1}F_{t+\ell+2}(\xb)F_{t+1}(\x)}{\x-\xb},
\end{align}
with $F_{\rho}(\x)={}_2F_1\left(\rho,\rho,2\rho;\x\right)$ being the standard hypergeometric function.

Due to operator mixing, the coefficients $A_{t,\ell}$ of the superconformal block decomposition are not in one-to-one correspondence with the OPE three-point functions $C_{ppK_i}$. Instead, they are given by a sum over the degenerate operators $K_i$:
\begin{align}
	A_{t,\ell} = \sum_{i=1}^{t-1}C_{ppK_i}C_{qqK_i}.
\end{align}
Similarly to the dimensions of exchanged operators $K_i$ (see eq.~(\ref{eq:dim_expansion})), we expand the three-point functions $C_{ppK_i}$ around large $N$ and $\l$,
\begin{align}
	C_{ppK_i} &= C^{(0)}_{ppK_i} + \left(\ll C^{(3)}_{ppK_i} +  \lll C^{(5)}_{ppK_i} + \ldots\right) + O(a),
	\label{eq:c_expansion}
\end{align}
where the superscript in $C_{ppK_i}^{(n)}$ again denotes the correction at order $\l^{-\frac{n}{2}}$.

The unmixing equations are stated most conveniently in a matrix form. Hence, let us assemble the three-point functions and anomalous dimensions at a given half-twist $t$ into the $(t-1)\times (t-1)$ matrices
\begin{align}
	\mathbb{C}^{(n)}:=\begin{pmatrix}C_{22K_1}^{(n)}&C_{22K_2}^{(n)}&\cdots&C_{22K_{t-1}}^{(n)}\\C_{33K_1}^{(n)}&C_{33K_2}^{(n)}&\cdots&C_{33K_{t-1}}^{(n)}\\\vdots&\vdots&&\vdots\\C_{ttK_1}^{(n)}&C_{ttK_2}^{(n)}&\cdots&C_{ttK_{t-1}}^{(n)}\end{pmatrix},\qquad\hat{\eta}^{(n)}:=\begin{pmatrix}\eta_1^{(n)}&&\\&\ddots&\\&&\eta_{t-1}^{(n)}\\\end{pmatrix},
\end{align}
with $\hat{\eta}^{(n)}$ being diagonal. We also arrange the correlators $\langle ppqq \rangle$ into the symmetric matrix
\begin{align}
	\hat{\mathcal{H}}(u,v):=\begin{pmatrix}\mathcal{H}_{2222}&\mathcal{H}_{2233}&\cdots&\mathcal{H}_{22tt}\\\mathcal{H}_{2233}&\mathcal{H}_{3333}&\cdots&\mathcal{H}_{33tt}\\\vdots&\vdots&&\vdots\\\mathcal{H}_{22tt}&\mathcal{H}_{33tt}&\cdots&\mathcal{H}_{tttt}\end{pmatrix}.
\end{align}
Now, plugging the double expansions (\ref{eq:dim_expansion}) and (\ref{eq:c_expansion}) into the superconformal block decomposition~(\ref{eq:block_deco}), we arrive at the decomposition
\begin{align}
\begin{split}
	\hat{\mathcal{H}}(u,v) = \sum_{t,\ell}\left[ \hat{A}^{(0,0)}_{t,\ell}+a\log(u)\left(\hat{A}^{(1,0)}_{t,\ell}+\ll\hat{A}^{(1,3)}_{t,\ell}+\ldots\right)+O(a^2)\right]G_{t,\ell}(u,v)+\ldots,
\end{split}
\end{align}
where the ellipsis denotes analytic terms in $u$ which are not relevant for this discussion.  Comparing to the expansion~(\ref{eq:H_expansion}) and keeping terms up to order $a\ll$, this leads to the unmixing equations
\begin{align}
	O(1):\qquad \hat{A}^{(0,0)}_{t,\ell}&=\mathbb{C}^{(0)}\left(\mathbb{C}^{(0)}\right)^T ,\\
	O(a):\qquad \hat{A}^{(1,0)}_{t,\ell}&=\mathbb{C}^{(0)}\hat{\eta}^{(0)}\left(\mathbb{C}^{(0)}\right)^T ,\\
	O(\ll):\hspace{1.56cm} 0&=\mathbb{C}^{(0)}\left(\mathbb{C}^{(3)}\right)^T+\mathbb{C}^{(3)}\left(\mathbb{C}^{(0)}\right)^T ,\label{eq:unmix_eq_3}\\
	O(a\ll):\qquad \hat{A}^{(1,3)}_{t,\ell}&=\mathbb{C}^{(0)}\hat{\eta}^{(3)}\left(\mathbb{C}^{(0)}\right)^T+\mathbb{C}^{(0)}\hat{\eta}^{(0)}\left(\mathbb{C}^{(3)}\right)^T+\mathbb{C}^{(3)}\hat{\eta}^{(0)}\left(\mathbb{C}^{(0)}\right)^T,\label{eq:unmix_eq_4}
\end{align}
where the zero on the lhs of equation~(\ref{eq:unmix_eq_3}) comes from the fact that are no $1/\l$ corrections to the leading $N$ free field correlator $\mathcal{H}^{(0,0)}$.

In~\cite{unmixing}, the first two equations were solved twist by twist to find an analytic formula for the supergravity anomalous dimensions $\hat{\eta}^{(0)}$ and to construct the leading order three-point function matrices $\mathbb{C}^{(0)}$.\footnote{Explicit data in the singlet channel is available up to twist 48. For some low twist examples see~\cite{unmixing}, where a straightforward prescription on how to compute the unmixed three-point functions in the singlet channel is given (for results in the $[0,1,0]$ channel see~\cite{Aprile:2017qoy}, respectively).} The conjectured anomalous dimension formula for a double-trace operator $\cO_{pq}$ in a general $su(4)$ channel $[a,b,a]$ reads~\cite{Aprile:2018efk}
\begin{align}
	\eta^{(0)}_{t,\ell}|_{[a,b,a]} =  -\frac{2M^{(4)}_t M^{(4)}_{t+\ell+1}}{\left(\ell+2(i+r)+a-\frac{1+(-1)^{a+\ell}}{2}\right)_6},
\label{eq:anom_dim_all_channels}
\end{align}
where 
\begin{align}
	M_t^{(4)} = (t-1)(t+a)(t+a+b+1)(t+2a+b+2),
\end{align}
the twist $\tau$ is parametrised by $t=\frac{\tau-b}{2}-a$ and the degeneracy labels $(i,r)$ are given by
\begin{align}
	i=1,\ldots,(t-1), \qquad r=0,\ldots,(\mu-1), \qquad \mu=\left\{\begin{array}{ll}
	\bigl\lfloor{\frac{b+2}2}\bigr\rfloor \quad &a+l \text{ even,}\\[.2cm]
	\bigl\lfloor{\frac{b+1}2}\bigr\rfloor \quad &a+l \text{ odd.}
	\end{array}\right.
\end{align}
Note that for $\mu>1$ and $t>2$ there are cases for which the labels $(i,r)$ assume the same sum $(i+r)$, resulting in a residual degeneracy in the supergravity spectrum.\footnote{\label{footnote:residual_degeneracy}The first instance of residual degeneracy occurs in the $[0,2,0]$ channel at twist 8 ($t=3$), where the two operators with labels $(i,r)=(1,1)$ and $(2,0)$ have the same supergravity anomalous dimension.} Furthermore, the supergravity anomalous dimension~\eqref{eq:anom_dim_all_channels} is left invariant under the discrete shift
\begin{align}
	t\rightarrow-t-\ell-2a-b-2,
\label{eq:sugra_symmetry}
\end{align}
which exchanges the two factors in its numerator. As we will discuss later, we find this symmetry to be present also in the $\ll$ and $\lll$ anomalous dimensions. 

With $\mathbb{C}^{(0)}$ and $\hat{\eta}^{(0)}$ at hand, we can turn our attention to the next two equations, where we have $\mathbb{C}^{(3)}$ and $\hat{\eta}^{(3)}$ as our unknowns. For a fixed half-twist $t$, we can first solve equation~(\ref{eq:unmix_eq_3}), leaving $\frac{(t-1)(t-2)}{2}$ undetermined parameters in $\mathbb{C}^{(3)}$. Together with the $t-1$ anomalous dimensions $\eta^{(3)}_i$ we have $\frac{t(t-1)}{2}$ unknowns, exactly as many as the number of conditions we get from $\hat{A}^{(1,3)}_{t,\ell}$ in equation~(\ref{eq:unmix_eq_4}). Thus this is a well defined problem with a unique solution.

In the following, we apply this analysis to the $[0,0,0]$ and $[0,1,0]$ channel, for which we use the $\langle ppqq \rangle$ and $\langle p(p+1)q(q+1) \rangle$ families of correlators. Note that our result~(\ref{eq:lambda32generalformula}) for the general four-point amplitude at $\ll$ provides the necessary data needed to resolve the mixing of double-trace operators in any $su(4)$ channel of the form $[a,b,a]$. We will address this general unmixing problem in the future.
\subsection{$[0,0,0]$ channel results}\label{sec:lambda_3/2_000}
Here we describe the solution of the full unmixing problem in the singlet channel. We obtain the necessary data, namely the correlators of the form $\langle ppqq \rangle$ at order $\ll$, from our new result~(\ref{eq:lambda32generalformula}). As expected, the conformal block decomposition yields only spin 0 contributions. Solving the unmixing equations~(\ref{eq:unmix_eq_3}) and ~(\ref{eq:unmix_eq_4}) twist by twist as described in the previous section, we find
\begin{align}
	\hat{\eta}^{(3)}=\left\{\eta^{(3)}_1,0,\ldots,0\right\},\qquad \mathbb{C}^{(3)}=0,
	\label{eq:unmixing_lambda32_000}
\end{align}
with $\eta^{(3)}_1$ being consistent with the formula
\begin{align}
	\eta^{(3)}_1 = -\frac{\zeta_3}{840} (t-1)^2 t^3 (t+1)^4(t+2)^3(t+3)^2\cdot\delta_{\ell,0}.
	\label{eq:anom_dim_lambda32_000}
\end{align}
Some comments are in order:
\begin{itemize}
	\item The leading $1/\l$ correction to the matrix of three-point functions $\mathbb{C}^{(3)}$ is identically zero. A priori, such a correction is not forbidden by consistency of the OPE and its vanishing is a very non-trivial result.\footnote{We would like to thank Ofer Aharony and Shai Chester for encouraging us to revisit the possibility of $1/\l$ corrections to the three-point functions.} We will use this observation and assume the absence of the subsequent correction $\mathbb{C}^{(5)}$ when exploring the spectrum of anomalous dimensions at order $\lll$ in the next section.
	\item The pattern of anomalous dimensions turns out to be surprisingly simple: only the operator with degeneracy label $i=1$ receives a $\ll$ correction to its dimension, all other anomalous dimensions vanish. As we will describe in section~\ref{sec:10d}, this pattern is consistent with predictions from the recently discovered hidden 10d conformal symmetry of supergravity correlators~\cite{Caron-Huot:2018kta}.
	\item For large $t$, the anomalous dimension has the asymptotic behaviour
	\begin{align}
		\eta_1^{(3)}\rightarrow -\frac{\zeta_3}{840}t^{14},
		\label{eq:large_twist_000}
	\end{align}
	a fact which will become important when comparing to the $[0,1,0]$ channel anomalous dimension. Furthermore, $\eta_1^{(3)}$ obeys the symmetry $t\rightarrow -t-2$.
	\item Lastly, our result for $\hat{\eta}^{(3)}$ correctly reproduces the averages of squared anomalous dimensions derived in~\cite{Alday:2018pdi}, see equation (3.11) therein.
\end{itemize}
\subsection{$[0,1,0]$ channel results}\label{sec:lambda_3/2_010}
Let us apply the method described above to solve the mixing problem in a non-trivial $su(4)$ channel. We choose the $[0,1,0]$ channel, where we need data from correlators of the form $\langle p(p+1)q(q+1) \rangle$. Again, we find contributions to spin 0 only. Performing the unmixing twist by twist we find similar results as in the singlet channel:
\begin{align}
	\hat{\eta}^{(3)}|_{[0,1,0]}=\left\{\eta^{(3)}_1|_{[0,1,0]},0,\ldots,0\right\},\qquad \mathbb{C}^{(3)}|_{[0,1,0]}=0,
	\label{eq:unmixing_lambda32_010}
\end{align}
where $\eta^{(3)}_1|_{[0,1,0]}$ is consistent with the formula
\begin{align}
	\eta^{(3)}_1|_{[0,1,0]} = -\frac{\zeta_3}{840} (t-1)^2 t^3 (t+1)^2 (t+2)^2 (t+3)^3 (t+4)^2\cdot\delta_{\ell,0},
	\label{eq:anom_dim_lambda32_010}
\end{align}
which is symmetric under $t\rightarrow -t-3$.

Again, the leading $1/\l$ correction to the matrix of three-point functions $\mathbb{C}^{(3)}|_{[0,1,0]}$ is zero and only the first operator with degeneracy label $i=1$ receives a  non-vanishing correction. For large $t$ it scales like the singlet channel anomalous dimension:
\begin{align}
	\eta_1^{(3)}|_{[0,1,0]}\rightarrow -\frac{\zeta_3}{840}t^{14}.
	\label{eq:large_twist_010}
\end{align}
This matching of the leading large twist behaviour of anomalous dimensions across different $su(4)$ channels is in agreement with the 10 dimensional interpretation discussed below.
\subsection{10d interpretation of spectrum}\label{sec:10d}
Recently, Caron-Huot and Trinh observed a hidden 10 dimensional conformal symmetry, which allowed them to repackage all half-BPS tree-level supergravity four-point functions into a single generating function~\cite{Caron-Huot:2018kta}.\footnote{See also~\cite{Rastelli:2019gtj} for a similar conjectured 6d version for AdS$_3\times S^3$ supergravity.} This conjecture was inspired by the following observation: the singlet channel supergravity anomalous dimension $\eta^{(0)}$ of a double-trace operator $\mathcal{O}_{pq}|_{[0,0,0]}$ and the partial-wave coefficients of the flat 10d $2\rightarrow2$ scattering amplitude of axi-dilatons in type IIB supergravity share a common Pochhammer structure in their denominators:
\begin{align}
	\frac{1}{\left(\ell_{10}+1\right)_6} \sim \frac{1}{\left(\ell+2i-1\right)_6},
\label{eq:10d_comparison_singlet_channel}
\end{align}
where the lhs depends on an effective 10 dimensional spin $\ell_{10}=0,2,\ldots$ and the rhs is the denominator of the supergravity singlet channel anomalous dimension $\eta^{(0)}|_{[0,0,0]}$, which depends on spin $\ell$ and degeneracy label $i$. For a general $su(4)$ channel $[a,b,a]$, we should compare to the denominator in equation~(\ref{eq:anom_dim_all_channels}), giving the correspondence
\begin{align}
	\frac{1}{\left(\ell_{10}+1\right)_6} \sim \frac{1}{\left(\ell+2(i+r)+a-\frac{1+(-1)^{a+\ell}}{2}\right)_6},
\label{eq:10d_comparison_all_channels}
\end{align}
where for convenience we repeat the definition of the degeneracy labels $(i,r)$:
\begin{align}
	i=1,\ldots,t-1, \qquad r=0,\ldots,\mu-1, \qquad \mu=\left\{\begin{array}{ll}
	\bigl\lfloor{\frac{b+2}2}\bigr\rfloor \quad &a+l \text{ even,}\\[.2cm]
	\bigl\lfloor{\frac{b+1}2}\bigr\rfloor \quad &a+l \text{ odd.}
	\end{array}\right.
\end{align}
The correspondence (\ref{eq:10d_comparison_all_channels}) assigns a value of the effective ten-dimensional spin to each long double-trace operator in the supergravity spectrum.

Now consider the first string correction at order $\ll$, the spectrum of which we have computed above for the singlet and $[0,1,0]$ channels, see equations~(\ref{eq:unmixing_lambda32_000}),~(\ref{eq:unmixing_lambda32_010}). Unexpectedly, we found that only the first anomalous dimension with degeneracy label $i=1$ is non-zero. A neat interpretation of this result can be given by the identification of denominators in~(\ref{eq:10d_comparison_all_channels}): as the order $\ll$ string correction descends from the 10d $\mathcal{R}^4$ supervertex, its ten-dimensional partial-wave decomposition contributes only to spin $\ell_{10}=0$. Matching denominators in equation~(\ref{eq:10d_comparison_singlet_channel}) we get
\begin{align}
	\ell_{10} = 0\quad \Rightarrow \quad (\ell,i) = (0,1),
	\label{eq:10d_prediction_spin0}
\end{align}
i.e. only the anomalous dimension with degeneracy label $i=1$ should be non-vanishing, which exactly coincides with our explicit results for $\eta^{(3)}$ in equations~(\ref{eq:unmixing_lambda32_000}) and~(\ref{eq:unmixing_lambda32_010})!

The heuristic correspondence~(\ref{eq:10d_comparison_all_channels}) thus seems to correctly give a prediction for which four-dimensional double-trace operators acquire an anomalous dimension, depending on the allowed ten-dimensional spin $\ell_{10}$. In fact, in the supergravity case, where operators of any even spin $\ell_{10}$ are exchanged, this does not result in any restrictions on the four-dimensional quantum numbers $(\ell,i,r)$, and indeed all operators are found to receive a correction to their dimensions. However, by assuming this ten-dimensional interpretation remains valid when considering further string corrections (which give only finite spin contributions to the partial-wave expansion), we can deduce constraints on the spectrum of anomalous dimensions, as shown above for the $\ll$ case.

The next string correction at order $\lll$ descends from the $\partial^4 \mathcal{R}^4$ supervertex, allowing spins up to $\ell_{10}=2$. Matching again the denominators, we find the prediction
\begin{align}
	\ell_{10} = 2\quad \Rightarrow \quad (\ell,i) = (2,1),~(1,1),~(0,2).
	\label{eq:10d_prediction_spin2}
\end{align}
Together with the spin $\ell_{10}=0$ contribution, we therefore expect three anomalous dimensions to be non-vanishing in the $[0,0,0]$ channel: the spin $2$, $i=1$ and the spin $0$, $i=1,2$ anomalous dimensions. In the $[0,1,0]$ channel, which also receives contributions from odd spins, we additionally expect the spin $1$, $i=1$ anomalous dimension to be present. 
As we will discuss in the next section, under an additional assumption about the three-point functions, the above structure can be used as a predictive tool to constrain further subleading corrections to the anomalous dimensions and correlators.

There is one further implication of the relation to ten dimensions, which concerns the observed coincidence of the large twist behaviour of anomalous dimensions, see equations~(\ref{eq:large_twist_000}) and~(\ref{eq:large_twist_010}) for a concrete example. 
In the 10 dimensional conformal four-point amplitude there is only one primary operator for any even spin $\ell_{10}$, which upon dimensional reduction results in multiple 4d primary operators descending from the same 10d primary.\footnote{We want to thank Paul Heslop for discussions on this point.}
We can use this to our advantage by making the following observation: 
For finite spin the large twist asymptotics accesses the flat space limit, which can be understood from the inverse Mellin transform, as defined in equation~\eqref{eq:mellintrafo}. The flat space limit tells us to look at the large $s,t$ behaviour, which in particular translates into large powers of $u$ in position space. Restricting ourselves to finite spin contributions, we then see that large twist indeed corresponds to the flat space limit. Schematically, for finite spin we thus have the correspondence
\begin{align}\label{eq:large_twist_asymptotics}
	\text{flat space limit} ~ \sim ~ \text{large twist asymptotics}.
\end{align}
In the flat space limit the spectrum is given by the 10d conformal theory. Therefore, at a given order in $1/\l$, we expect the same large twist asymptotics for all four dimensional operators which descend from a common 10d primary according to equation~(\ref{eq:10d_comparison_all_channels}), regardless of their 4d quantum numbers (such as spin, degeneracy labels or $su(4)$ representation). This cross-channel matching of the leading twist behaviour can serve as a consistency check on results for anomalous dimensions and will allow us to fix one more free parameter in the $\lll$ $[0,1,0]$ channel spectrum.

\section{New results at order $\lll$}\label{sec:unmixing_lambda_5/2}\setcounter{equation}{0}
In the previous section we have examined the spectrum of anomalous dimensions of double-trace operators in the singlet and the $[0,1,0]$ channels for the first $1/\lambda$ correction. Here, we will turn our attention to the next order, namely $\lll$, and apply the unmixing procedure in a similar fashion as before. We will first discuss the singlet channel, for which recently the $\langle 22pp \rangle$ family of correlators was completely determined up to this order~\cite{Binder:2019jwn, Alday:2018pdi}. In the last part of this section, we will consider the $[0,1,0]$ channel and present results for a new family of correlators of the form $\langle 23(p-1)p \rangle$.
\subsection{The singlet channel spectrum}
Let us start by adapting the general ansatz~\eqref{eq:mellin_expansions} to the $\langle 22pp \rangle$ family of correlators: 
\begin{align}
\me^{(1,5)}_{22pp}(s,t)=B^{6}_{6,1} \me^{6,1}_p +B^{6}_{6,2} \me^{6,2}_p + B^{6}_5 \me^5_p + B^{6}_4 \me^4_p,
\label{eq:lambda_5/2_mellin_amplitude}
\end{align}
where the basis of Mellin space amplitudes $\me^n_p(s,t)$ has one remaining crossing symmetry ($t\leftrightarrow u)$ and is given by~\cite{Alday:2014tsa}
\begin{align}
\begin{split}
	\me^4_p&=1,\qquad\qquad\qquad~~~\hspace{0.05cm}\me^5_p=s,\\
	\me^{6,1}_p&=s^2+t^2+u^2,\qquad\me^{6,2}_p=s^2,
\end{split}\label{eq:lambda_5/2_mellin_basis}
\end{align}
and their explicit position space results are given in Appendix \ref{app:dbars}.\footnote{Note that compared to the previous order (where $\me_p^4$ contributes to spin 0 only), we find an additional spin 2 contribution from $\me_p^{(6,1)}$. Hence, in order to distinguish the two spin contributions, we will use an additional subscript in the $\lll$ anomalous dimension $\eta^{(5)}_{\ell,i}$, with $i=1,\ldots,t-1$ as before and $\ell$ labelling the different spins. We will find contributions from $\ell=0,2$ in the singlet and from $\ell=0,1,2$ in the $[0,1,0]$ channel.}

The corresponding $p$-dependent singlet channel coefficients $B^k_n$ have been shown to obey the following form based on arguments of locality on $S^5$~\cite{Binder:2019jwn}:\footnote{The extra factor of $p$ compared to~\cite{Binder:2019jwn} is due to a different normalisation of the external operators, see equation~\eqref{eq:single_trace_op} for our conventions.}
\begin{align} \label{eq:poly form 22pp}
B^{k}_{n} = \frac{p~(p)_n}{(p-2)!} ~ C^{k}_{n}(p),
\end{align}
where $C^{k}_{n}(p)$ is a polynomial of degree $2(k-n)$ in $p$. It will turn out that a similar result also holds in the $[0,1,0]$ channel. These polynomials have been fully fixed in~\cite{Alday:2018pdi,Binder:2019jwn}, and are given by
\begin{align} \label{eq: 22pp polynomials}
\begin{aligned}
B^{6}_{4} &= -  \frac{p~(p)_{4}\zeta_{5} }{(p-2)!}~2(p^4 + 9p^3 + 10p^2-20p-25), &&& B^{6}_{5} &= \frac{p~(p)_{5}\zeta_{5}}{(p-2)!}~2p(p-2),\\
B^{6}_{6,1} &= \frac{p~(p)_{6}\zeta_{5}}{(p-2)!}, &&& B^{6}_{6,2} &= 0. 
\end{aligned}
\end{align}
In order to solve the full mixing problem (i.e. determining the $t-1$ anomalous dimensions $\eta^{(5)}_{\ell,i}$ as well as the matrix of three-point function corrections $\mathbb{C}^{(5)}$), we would need the full $\langle ppqq \rangle$ family of correlators at $\lll$. However, we can circumvent this obstacle by making an assumption about the corrections to the three-point functions, which is motivated by results at the previous order. Recall that by explicit computations in the singlet channel we have determined $\mathbb{C}^{(3)} = 0$. Let us make the analogous assumption here, namely
\begin{align}\label{eq: assumption 3pt functions}
	\mathbb{C}^{(5)} = 0.
\end{align}
We do not know if the above assumption is valid in general. However we will now show that the resulting spectrum of anomalous dimensions is in agreement with the 10d predictions~\eqref{eq:10d_prediction_spin2}. The assumption (\ref{eq: assumption 3pt functions}) then allows us to solve the mixing problem for the $\lll$ anomalous dimensions using data from the known $\langle 22pp \rangle$ series of correlators only:

For a given half-twist $t$ and spin $\ell$, the only unknowns are the anomalous dimensions $\eta^{(5)}_{\ell,i}$, and there are $t-1$ of them. At the same time, we get $t-1$ equations from considering the set of correlators $\left\{\langle 2222 \rangle,\langle 2233 \rangle,\ldots,\langle 22tt \rangle\right\}$. To be more concrete, let us denote the conformal block decomposition of the $\log(u)$ part of $\mathcal{H}^{(1,5)}$ for the correlator $\langle 22pp\rangle$ by $A^{(1,5)}_{p,\ell}$. Employing a matrix notation for the mentioned set of $t-1$ equations, we obtain
\begin{align}\label{eq:unmixing_equations_simplified}	\begin{pmatrix}C^{(0)}_{22K_1}C^{(0)}_{22K_1}&\cdots&C^{(0)}_{22K_{t-1}}C^{(0)}_{22K_{t-1}}\\\vdots&&\vdots\\C^{(0)}_{22K_1}C^{(0)}_{ttK_1}&\cdots&C^{(0)}_{22K_{t-1}}C^{(0)}_{ttK_{t-1}}\end{pmatrix}\begin{pmatrix}\eta^{(5)}_{\ell,1}\\\vdots\\\eta^{(5)}_{\ell,t-1}\end{pmatrix} = \begin{pmatrix}A^{(1,5)}_{2,\ell}\\\vdots\\A^{(1,5)}_{t,\ell}\end{pmatrix}.
\end{align}
Since the matrix on the lhs is known explicitly~\cite{unmixing}, we can invert the above equation for a given $t$ and readily obtain the vector of anomalous dimensions $\hat{\eta}_{\ell}^{(5)}$.

Let us move on to the concrete unmixing of the singlet channel, where we start by considering the spin 2 contribution. Using the procedure described above, we find that the only spin 2 contribution comes from the combination $u^p(1+v)\dbar{p+3p+355}$, and the $\ell=2$ anomalous dimensions $\hat{\eta}^{(5)}_2$ follow the pattern predicted in equation~($\ref{eq:10d_prediction_spin2}$), i.e. only the $i=1$ anomalous dimension is non-vanishing:
\begin{align}
	\hat{\eta}^{(5)}_{2}=\left\{\eta^{(5)}_{2,1},0,\ldots,0\right\}, 
	\label{eq:unmixing_lambda52_000_spin2}
\end{align}
with $\eta^{(5)}_{2,1}$ being consistent with the formula 
\begin{align}
	\eta^{(5)}_{2,1} = -\frac{\zeta_5}{166320}(t-1)^2 t^2 (t+1)^3 (t+2)^4 (t+3)^3 (t+4)^2 (t+5)^2\cdot\delta_{\ell,2},
	\label{eq:anom_dim_lambda52_000_spin2}
\end{align}
which is symmetric under $t\rightarrow -t-4$. In the large $t$ limit we find the asymptotic behaviour 
\begin{align}
\eta^{(5)}_{2,1} \rightarrow -\frac{\zeta_5}{166320}t^{18}.
\end{align}
Moving on to the spin 0 contributions, the combinations $D^{4}_p,D^{5}_p,D^{6,1}_p, D^{6,2}_p$ described in equation~\eqref{eq:dbars 22pp} all contribute. Solving the unmixing equation~\eqref{eq:unmixing_equations_simplified} we find that only the operators with degeneracy labels $i=1,2$ receive a correction, which is again in total agreement with our $10$d interpretation of the spectrum. We obtain
\begin{align}
	\hat{\eta}^{(5)}_{0}=\left\{\eta^{(5)}_{0,1},\eta^{(5)}_{0,2},0,\ldots,0\right\},
	\label{eq:unmixing_lambda52_000_spin0}
\end{align}
with $\eta^{(5)}_{0,1},\eta^{(5)}_{0,2}$ being consistent with the formulae
\begin{align} \label{eq:anom_dim_lambda52_000_spin0}
\begin{aligned}
	\eta^{(5)}_{0,1} &= -\zeta_5 ~ \frac{(t-1)^2 t^3 (t+1)^4 (t+2)^3 (t+3)^2 \left(4 t^4+16 t^3+14 t^2-4 t+15\right)}{4320}  \cdot\delta_{\ell,0}, \\
\eta^{(5)}_{0,2} &= -\zeta_5 ~ \frac{(t-1)^2 t^3 (t+1)^4 (t+2)^3 (t+3)^2 \left(t^4+4 t^3+26 t^2+44 t-30\right)}{166320} \cdot\delta_{\ell,0}
\end{aligned}
\end{align}
which both have the symmetry $t \rightarrow -t-2$.  In the large $t$ limit they behave as
\begin{align}
\begin{aligned}
\eta^{(5)}_{0,1} &\rightarrow -\frac{\zeta_5}{1080}t^{18}, \\
\eta^{(5)}_{0,2} &\rightarrow -\frac{\zeta_5}{166320}t^{18},
\end{aligned}
\end{align}
and we note that $\eta^{(5)}_{0,2}$ scales precisely the same way as $\eta^{(5)}_{2,1}$, as expected from our $10$d interpretation~($\ref{eq:10d_prediction_spin2}$), since they descend from the same $10$d spin $\ell_{10}=2$ operator. 
\subsection{Spectrum constraints} \label{sec: spectrum constraint}
We have already seen that various techniques constrain the form of the $p$-dependent polynomials and we have also observed very nice and remarkably simple structures in the spectra of operators at orders $\ll$ and $\lll$. Let us now discuss an alternative approach.

We will assume that we do not know the exact form of the $p$-dependent coefficients $B^6_n$ and we write general polynomial ans{\"a}tze.
Before proceeding, there are some constraints that we can take into account already at this stage. Firstly, the coefficient $B^{6}_{6,2} $ has to vanish due to the absence of the corresponding term in the flat space limit. Moreover, for $p=2$ the coefficient $B^{6}_{5}$ has to vanish as a consequence of crossing symmetry. The ans{\"a}tze we make for the coefficients $B^{6}_{n}$ are then: 
\begin{align} \label{eq: 22pp poly ansatze}
\begin{aligned}
B^{6}_{4} &= \zeta_5 ~ \frac{p}{(p-2)!}\sum_{i=0}^{9} \beta_{i} p^i, &&& B^{6}_{5} &=  \zeta_5 ~ \frac{p}{(p-2)!} ~ (p-2) ~ \sum_{i=0}^{7} \gamma_{i}  p^i, \\
B^{6}_{6,1} &= \zeta_5 ~ \frac{p}{(p-2)!}\sum_{i=0}^{7} \delta_{i} p^i, &&& B^{6}_{6,2} &= 0.
\end{aligned}
\end{align}
Let us point out once more that we are working under the assumption that the string corrections to the three point functions are vanishing at this order, see eq.~($\ref{eq: assumption 3pt functions}$).

We start by examining the spin $2$ sector, where only the coefficient $B^6_{6,1}$ contributes. Solving the reduced unmixing equation~\eqref{eq:unmixing_equations_simplified}, finding that all of the operators receive a non-vanishing anomalous dimensions, which depend on the free parameters introduced by $B^6_{6,1}$. According to the 10d prediction described in section~\ref{sec:10d}, we now impose that all but the $i=1$ anomalous dimension are zero, i.e. we demand that $\eta_{2,i}^{(5)} = 0$, for $i=2,\ldots,t-1$. Imposing these constraints on $B^6_{6,1}$ we find
\begin{align}
B^6_{6,1} = \zeta_5 \frac{p~(p)_6}{120 (p-2)!}~ \delta_{1},
\end{align}
which is the correct form of the polynomial up to a free parameter. The formula for the unmixed anomalous dimension is given by eq.~($\ref{eq:anom_dim_lambda52_000_spin2}$) up to the free parameter $\delta_1$, which can be fixed by matching to the flat space limit (determining $\delta_1=120$).

Having completely fixed one more polynomial out of the four in eq. ($\ref{eq: 22pp poly ansatze}$), we proceed to the spin $0$ sector. As before, we unmix the spectrum of operators and we observe that all of them receive a non-zero anomalous dimension. Following the 10d prescription, we impose the vanishing of all anomalous dimensions $\eta_{0,i}^{(5)}$ with $i>2$, obtaining
\begin{align}
\begin{aligned}
B^{6}_{4} &= - \zeta_5 ~ \frac{p~(p)_4}{(p-2)!} P_{4}(p),\\
B^{6}_{5} &= - \zeta_5 ~ \frac{p~(p)_5}{(p-2)!} ~ \frac{(p-2)}{576}\big((288+11\beta_1-6\beta_2)p+2(1440+11\beta_1-6\beta_2)\big),
\end{aligned}
\end{align} 
where $P_4(p)$ is the fourth-order polynomial
\begin{align}
\begin{split}
P_{4}(p) &= \frac{1}{432}\big(864p^4+3(1152-11\beta_1+6\beta_2)p^3-2(85\beta_1-66\beta_2+36\beta_3)p^2\\
&\qquad\quad~~+12(11\beta_1-6\beta_2)p-72\beta_1\big).
\end{split}
\end{align}
We have obtained the predicted form of the coefficients $B^6_n$, see equation~($\ref{eq:poly form 22pp}$), by using our spectrum constraints, crossing symmetry and the flat-space limit, under the assumption of a guided polynomial ansatz. The observation about the form of the polynomials will be very useful in the study of the $[0,1,0]$ channel, as we discuss in the next section.

In fact, we are able to fix one more free parameter by comparing the large $t$ limit of the anomalous dimensions. The leading $t$ power of the spin 0 anomalous dimension $\eta_{0,2}^{(5)}$ depends on two free parameters, namely $\beta_1$ and $\beta_2$. Matching this to the large $t$ limit of $\eta_{2,1}^{(5)}$, which descends form the same 10d operator with $\ell_{10}=2$, determines
\begin{align}
\beta_2 = \frac{11}{6} \beta_1 +240 \,.
\end{align}
The above condition yields
\begin{align}
\begin{aligned}
B^{6}_{4} &= - \zeta_5 ~ \frac{p~(p)_4}{(p-2)!} \Bigl(2 p^4+18 p^3+\frac{220 p^2}{3}-40 p  + \frac{\beta_1}{6}( p^2 -1) -\frac{\beta_3}{6}  p^2 \Bigr),\\
B^{6}_{5} &= \zeta_5 ~ \frac{p~(p)_5}{(p-2)!} ~ 2 p (p-2)\,,
\end{aligned}
\label{finalform}
\end{align} 
which is equivalent to the result of \cite{Alday:2018pdi} up to a redefinition of the two remaining free parameters.

The localisation conditions of \cite{Binder:2019jwn} also fix $B^{6}_{4}$ and $B^{6}_{5}$ up to two free parameters. As observed in \cite{Binder:2019jwn}, combining the localisation constraints with the results of \cite{Alday:2018pdi}, or equivalently the form (\ref{finalform}), completely fixes the Mellin amplitude. 
If we take the localisation constraints of \cite{Binder:2019jwn} but do not combine them with the results of \cite{Alday:2018pdi} then the spectrum obtained is such that all operators receive a non-vanishing anomalous dimension. Imposing the vanishing of the anomalous dimensions of operators of ten-dimensional spin $\ell_{10}>2$ then also fixes the final form of the Mellin amplitude (\ref{eq: 22pp polynomials}).
\subsection{The $\langle 23(p-1)p\rangle$ family of correlators and the $[0,1,0]$ spectrum} \label{sec:010}
We have seen in the previous section that imposing the spectrum constraints, together with matching the large $t$ asymptotics of the anomalous dimensions according to our 10d prescription, allows us to derive the correct $p$-dependent coefficients $B^6_n$ in the singlet channel up to two free parameters. Here, we will use the same approach to study for the first time the $\lll$ correction to the $\langle 23(p-1)p \rangle$ family of correlators, from which we can extract information about the $[0,1,0]$ channel anomalous dimensions, denoted by $\eta_{\ell,i}^{(5)}|_{[0,1,0]}$ in the following. We parametrise this family by the same $p$ as in $\langle 22pp \rangle$, such that for $p=3$ the two families coincide up to a crossing transformation.

We apply the crossing transformation exchanging the points $2\leftrightarrow3$ (in Mellin space, this amounts to swapping $s\leftrightarrow u$) to the Mellin space basis functions given in equation~\eqref{eq:lambda_5/2_mellin_basis}, we find
\begin{align}
\begin{split}
	\me^{4}_p|_{[0,1,0]}&=1,\qquad~\me^{5,1}_p|_{[0,1,0]}=s,\\
	\me^{5,2}_p|_{[0,1,0]}&=u,\qquad~\me^{6,1}_p|_{[0,1,0]}=s^2+t^2+u^2,\\
	\me^{6,2}_p|_{[0,1,0]}&=s^2,\qquad\me^{6,3}_p|_{[0,1,0]}=u^2,
\end{split}\label{eq:lambda_5/2_mellin_basis_010}
\end{align}
where we had to add the two additional independent polynomial basis elements $\me^{5,2}_p|_{[0,1,0]}$ and $\me^{6,3}_p|_{[0,1,0]}$, because the correlator $\langle23(p-1)p\rangle$ has no remaining crossing symmetry. Their explicit position space results are given in Appendix \ref{app:dbars}, from which we find finite spin contributions to spins $\ell=0,1,2$ only.

For the $p$-dependent coefficients $B^6_{n,i}|_{[0,1,0]}$ associated with the above Mellin amplitudes we make the ans\"atze
\begin{align}
\begin{aligned}
B^6_4|_{[0,1,0]} &=\frac{1}{(p-3)!}\sum_{i=0}^9 \widetilde{\beta}_{i} p^i, &&& B^6_{5,1}|_{[0,1,0]} &=\frac{1}{(p-3)!}\sum_{i=0}^8 \widetilde{\gamma}_{i} p^i,\\
B^6_{5,2}|_{[0,1,0]}&=\frac{1}{(p-3)!}\sum_{i=0}^8 \widetilde{\delta}_{i} p^i, &&& B^6_{6,1}|_{[0,1,0]}&= \zeta_5 ~ \frac{3(p-1)_7}{2(p-3)!},\\
B^6_{6,2}|_{[0,1,0]}&=0, &&& B^6_{6,3}|_{[0,1,0]}&=0,
\end{aligned}\label{eq:coeffs_ansatz_010}
\end{align}
where the vanishing of $B^6_{6,2}|_{[0,1,0]}$ and $B^6_{6,3}|_{[0,1,0]}$, and the form of $B^6_{6,1}|_{[0,1,0]}$ follow from matching against the flat space limit.

As in the singlet channel, we make the assumption of vanishing $\lll$ corrections to the three-point functions, namely
\begin{align} \label{eq: assumption 3pt functions 010}
\mathbb{C}^{(5)}|_{[0,1,0]}=0.
\end{align}
Under this assumption, we can use the Mellin amplitudes~\eqref{eq:lambda_5/2_mellin_basis_010} together with the coefficients~\eqref{eq:coeffs_ansatz_010} to unmix the spectrum of $[0,1,0]$ channel anomalous dimensions $\eta^{(5)}_{\ell,i}|_{[0,1,0]}$ and obtain the following:
\begin{itemize}
\item The spin $2$ operators receive a contribution from $B^6_{6,1}|_{[0,1,0]}$ only and the spectrum turns out to be of the same form as in the singlet channel discussed previously, i.e. only the operators with $i=1$ receive a correction, confirming our $10$d prescription in a non-trivial $su(4)$ channel.

\item In the spin $1$ sector, we have an additional contribution from  $B^6_{5,2}|_{[0,1,0]}$. Unmixing the spectrum and imposing the vanishing of all anomalous dimensions with $i>1$, we are left with one undetermined parameter from $B^6_{5,2}|_{[0,1,0]}$. We can fix its value by matching the large $t$ limit of the spin 1 anomalous dimension against the spin 2 scaling, as they descend from the same 10d spin.

\item Finally, we proceed by unmixing the spin $0$ sector, to which all coefficients $B^6_n|_{[0,1,0]}$ contribute. We are using the $10$d interpretation as a prediction mechanism and we solve the spectrum constraint by imposing zeroes on the anomalous dimensions appropriately, see eq. ($\ref{eq:10d_prediction_spin0}$). Imposing these spectrum conditions, we are left with four free parameters. Both anomalous dimensions depend on these four parameters and have the expected large $t$ behaviour, $\eta_{0,i}^{(5)}|_{[0,1,0]}\rightarrow t^{18}$.\\
For $p=3$, we can match the coefficient polynomials to the $\langle 2323\rangle$ result, which we obtain by crossing from the known $\langle 2233 \rangle$ correlator. This provides us with an additional constraint for each of the non-vanishing polynomials, reducing the number of undetermined parameters down to two. Furthermore, we use the large $t$ limit to perform a cross-channel match for the $i=1$ spin 0 anomalous dimension: $\eta_{0,1}^{(5)}|_{[0,0,0]} \sim \eta_{0,1}^{(5)}|_{[0,1,0]}$ for large $t$, which fixes one more parameter. Therefore, we are finally left with one undetermined parameter: $\widetilde{\beta}_{1}$.
\end{itemize}
Imposing all of the above constraints, the coefficients $B^6_{n}|_{[0,1,0]}$ turn out to be
\begin{align}\label{eq:B_coeffs_010}
\begin{aligned}
B^6_4|_{[0,1,0]} &=\frac{(p-1)_5\zeta_5}{36(p-3)!}~P(p),&&& B^6_{5,1}|_{[0,1,0]} &=\frac{3(p-1)_6 \zeta_5 }{(p-3)!}~p(p-3),\\
B^6_{5,2}|_{[0,1,0]}&=\frac{3(p-1)_6\zeta_5}{(p-3)!}~p, &&& B^6_{6,1}|_{[0,1,0]}&= \frac{3(p-1)_7\zeta_5 }{2(p-3)!},\\
B^6_{6,2}|_{[0,1,0]}&=0, &&& B^6_{6,3}|_{[0,1,0]}&=0,
\end{aligned}
\end{align}
with $P(p)$ given by
\begin{align}
	P(p)= - 18 p \left(6 p^3+60 p^2+41 p-167\right) + (p-3) (p+2)\widetilde{\beta}_{1}.
\end{align}
Note that the above coefficients are consistent with the additional crossing symmetry of the $\langle2334\rangle$ correlator: at $p=4$ we find that $B^6_{5,1}|_{[0,1,0]}=B^6_{5,2}|_{[0,1,0]}$, which we did not use as an input and hence serves as a consistency check.

We observe that in this case, similarly to the singlet channel formula~\eqref{eq:poly form 22pp}, the coefficients $B^{k}_{n}|_{[0,1,0]}$ are consistent with the general form 
\begin{align}
	B^k_n|_{[0,1,0]}  = \frac{(p-1)_{n+1}}{(p-3)!}C^k_n,
\end{align}
where $C^k_n$ is a polynomial in $p$ of degree $2(k-n)$. This is the analogous result of locality on $S^5$ for the $\langle 23(p-1)p\rangle$ correlators considered here.

For the $[0,1,0]$ anomalous dimensions, the constraints described above give results consistent with
\begin{align}
\begin{aligned}
	\eta^{(5)}_{2,1}|_{[0,1,0]} &= - \frac{\zeta_5}{166320} (t-1)^2 t^2 (t+1) (t+2)^4 (t+3)^4 (t+4) (t+5)^2 (t+6)^2 \cdot \delta_{\ell,2}, \\
	\eta^{(5)}_{1,1}|_{[0,1,0]} &= - \frac{\zeta_5}{332640} (t-1)^2 t^2 (t+1)^2 (t+2)^4 (t+3)^2 (t+4)^2 (t+5)^2 \left(2 t^2+8 t+1\right) \cdot \delta_{\ell,1}, \\
	 \eta^{(5)}_{0,1}|_{[0,1,0]}  &= -\frac{\zeta_5}{3265920} (t-1)^2 t^3 (t+1)^2 (t+2)^2 (t+3)^3 (t+4)^2  \cdot Q_1(t) \cdot \delta_{\ell,0}, \\
	\eta^{(5)}_{0,2}|_{[0,1,0]}  &= -\frac{\zeta_5}{17962560} (t-1)^2 t^3 (t+1)^2 (t+2)^2 (t+3)^3 (t+4)^2 \cdot Q_2(t)  \cdot \delta_{\ell,0},
\end{aligned}
\end{align}
where $Q_1(t)$ and $Q_2(t)$ are the fourth order polynomials
\begin{align}
\begin{split}
	Q_1(t) &=18 \left(168 t^4+1008 t^3+1403 t^2-327 t-380\right)+5 (t-2) (t+5)\widetilde{\beta}_{1},\\
	Q_2(t)  &=18 \left(6 t^4+36 t^3+206 t^2+456 t+205\right) -(2 t+1) (2 t+5)\widetilde{\beta}_{1}.
\end{split}
\end{align}
Let us at this point make a remark on the symmetries of the anomalous dimensions: firstly, $\eta^{(5)}_{2,1}|_{[0,1,0]}$ is symmetric under $t\rightarrow -t-5$. The spin 1 anomalous dimension, $\eta^{(5)}_{1,1}|_{[0,1,0]}$, is symmetric under $t\rightarrow -t-4$ and finally the spin 0 formulae have a symmetry under $t \rightarrow -t-3$, which is in complete agreement with the symmetry the supergravity anomalous dimension was found to obey, see equation~\eqref{eq:sugra_symmetry}. 

We end this section by commenting on the validity of our results for the $\langle 23(p-1)p\rangle$ family of correlators given in~\eqref{eq:B_coeffs_010}. In the singlet channel, the coefficients $B^6_k$ for the correlators $\langle 22pp \rangle$ were derived using constraints obtained by matching the tree-level and one-loop flat space amplitudes~\cite{Alday:2018pdi} and localisation~\cite{Binder:2019jwn}, respectively. In contrast, in the $[0,1,0]$ channel, for which no localisation constraints currently exist, we used three different sets of constraints: matching against the tree-level flat space amplitude, matching known results at $p=3$ as well as our spectrum constraints, which are based on the assumption that $\mathbb{C}^{(5)}|_{[0,1,0]} = 0$. Even though we can not give a proof of this assumption, we believe that our final results for the $\langle 23(p-1)p \rangle$ correlators are correct in any case, as they are independent of the precise form of the spectrum. In the case that $\mathbb{C}^{(5)}|_{[0,1,0]}$ is non-zero, we expect a modification in the subleading large $t$ behaviour of the anomalous dimensions $\eta_{\ell,i}^{(5)}|_{[0,1,0]}$. However, the leading large $t$ asymptotics would not change because it is fully determined by the flat space limit, as argued above equation~\eqref{eq:large_twist_asymptotics}.
\section{Conclusions}\setcounter{equation}{0}
Let us conclude with mentioning some open questions and possible future directions:
\begin{itemize}
\item We have seen that the first $1/\l$ correction to the three-point functions in the singlet and $[0,1,0]$ channel vanish, i.e. $\mathbb{C}^{(3)}|_{[0,0,0]}=\mathbb{C}^{(3)}|_{[0,1,0]}=0$. We expect this result to extend to all $su(4)$ channels. These careful cancellations in the unmixing equations are highly non-trivial and it would be fascinating to get a better understanding of this surprising result from a purely CFT point of view. It would also be interesting to test our working assumption of the vanishing at higher orders in $1/\lambda$.

\item As already mentioned before, our result~\eqref{eq:lambda32generalformula} for the order $\ll$ correlator with arbitrary external charges provides the necessary data to attack the general mixing problem for any $su(4)$ channel $[a,b,a]$. This general analysis can be carried out along the lines of~\cite{Aprile:2018efk}, where the corresponding supergravity mixing problem was resolved. Using our 10 dimensional interpretation of the spectrum, we expect only the spin~0 anomalous dimension in the $[0,b,0]$ channel with degeneracy labels $(i,r)=(1,0)$ to be non-vanishing. We hope to report on this in the near future.

\item We would like to consider higher $1/\l$ corrections and investigate how constraining our spectrum conditions are. In particular, it would be interesting to combine the spectrum conditions with constraints obtained from extending other methods to higher orders, for example results from supersymmetric localisation~\cite{Binder:2019jwn} or matching with one-loop string amplitudes in the bulk-point limit~\cite{Alday:2018pdi}.

\item We believe that the full implications of the observed 10 dimensional conformal symmetry of supergravity amplitudes~\cite{Caron-Huot:2018kta} are still not fully explored. The observed pattern in the string corrections to the spectrum hints at their common 10 dimensional origin, which we conjecture to persist to all orders in $1/\l$. An additional hint comes from considering the spectrum in a large twist limit, where anomalous dimensions descending from the same 10d operator share the same leading large twist asymptotics. It would be very interesting to investigate this connection further.

\item We observed that the $\ll$ and $\lll$ anomalous dimensions are invariant under the discrete symmetry $t\rightarrow-t-\ell-2a-b-2$, a symmetry which is also present in the supergravity anomalous dimensions $\eta^{(0)}$. We would like to gain a better understanding of the origin and implications of this symmetry.
\end{itemize}
\section*{Acknowledgements}
JMD, DN and HP acknowledge support from ERC Consolidator grant 648630 IQFT. We would like to thank Ofer Aharony, Shai Chester and Paul Heslop for useful discussions.

\appendix
\section{$\dbar{}$-representation of Mellin basis}\setcounter{equation}{0}
\label{app:dbars}
Here we collect the position space results $D^n_p$ corresponding to the singlet and $[0,1,0]$ channel Mellin amplitudes $\me^n_p$ associated with the two families of correlators we study. For the singlet channel, we find have
\begin{align} \label{eq:dbars 22pp}
\begin{split}
	D^4_p &= u^p\dbar{p+2,p+2,4,4}(u,v),\\
	D^5_p &= 2u^p\left(2\dbar{p+2,p+2,4,4}(u,v)-\dbar{p+2,p+2,5,5}(u,v)\right),\\
	D^{6,1}_p &= 2u^p\left(2(1+u+v)\dbar{p+3,p+3,5,5}(u,v)-(4+4p-p^2)\dbar{p+2,p+2,4,4}(u,v)\right),\\
	D^{6,2}_p &= 4u^p\left(\dbar{p+2,p+2,6,6}(u,v)-5\dbar{p+2,p+2,5,5}(u,v)+4\dbar{p+2,p+2,4,4}(u,v)\right),
\end{split}
\end{align}
where we used various identities amongst $\dbar{}$-functions to simplify the results.\footnote{See for example~\cite{Arutyunov:2002fh} for a useful collection of $\dbar{}$-identities.}

As expected, we find that the $D^{n}_{p}$ with $n>1$, corresponding to polynomial Mellin amplitudes, give only finite spin contributions. Note that the $\dbar{}$-functions appearing in~\eqref{eq:dbars 22pp} can be classified according to their highest spin contributions, and we find they fall into the three families 
\begin{align} \label{eq:experimental spins}
\begin{split}
	u^p\dbar{p+2,p+2,n,n}\qquad&\rightarrow\qquad\text{spin 0},\\
	u^{p+1}\dbar{p+3,p+3,n,n}\qquad&\rightarrow\qquad\text{spin 0},\\
	(1+v)u^p\dbar{p+3,p+3,n,n}\qquad&\rightarrow\qquad\text{spin 0,2}.
\end{split}
\end{align}
For the position space results $D^n_p|_{[0,1,0]}$ corresponding to the $[0,1,0]$ channel Mellin amplitudes $\me_p^n|_{[0,1,0]}$ we obtain
\begin{align}
\begin{split}
	D^4_p|_{[0,1,0]} &= u^{p-1}\dbar{p+1,p+2,4,5}(u,v),\\
	D^{5,1}_p|_{[0,1,0]} &= u^{p-1} \left(5\dbar{p+1,p+2,4,5}(u,v)-2\dbar{p+1,p+2,5,6}(u,v)\right),\\
	D^{5,2}_p|_{[0,1,0]} &= u^{p-1} \left( (p - 7) \dbar{p+1,p+2,4,5}(u,v) +2 \dbar{p+1,p+2,5,6}(u,v) + 2 \dbar{p+2,p+2,4,6}(u,v) \right),\\
	D^{6,1}_p|_{[0,1,0]} &= 2u^{p-1} \left(\left(p^2 -5 p +39\right)\dbar{p+1,p+2,4,5}(u,v) + 2 (p-14) \dbar{p+1,p+2,5,6}(u,v) \right.\\ &~~~~\left.+ 4 \dbar{p+1,p+2,6,7}(u,v)-22 \dbar{p+2,p+2,4,6}(u,v) + 4 \dbar{p+2,p+2,5,7}(u,v) + 4 \dbar{p+3,p+2,4,7}(u,v) \right),\\
	D^{6,2}_p|_{[0,1,0]} &= u^{p-1} \left( 25 \dbar{p+1,p+2,4,5}(u,v) - 24 \dbar{p+1,p+2,5,6}(u,v) + 4 \dbar{p+1,p+2,6,7}\right),\\
	D^{6,3}_p|_{[0,1,0]} &= u^{p-1} \left(\left(p-7\right)^2 \dbar{p+1,p+2,4,5}(u,v) + 4(p-8) \dbar{p+1,p+2,5,6}(u,v) + 4\dbar{p+1,p+2,6,7} \right.\\ &~~~~\left. + 4 (p-8) \dbar{p+2,p+2,4,6}(u,v) + 8 \dbar{p+2,p+2,5,7}(u,v) + 4\dbar{p+3,p+2,4,7} \right).
\end{split}
\end{align}
As in the singlet channel we find that we have finite spin contributions only, and we observe that the above $\dbar{}$-functions fall into the three families:
\begin{align} \label{eq:experimental spins part II}
\begin{split}
	u^{p-1}\dbar{p+1,p+2,n,n+1}\qquad&\rightarrow\qquad\text{spin 0},\\
	u^{p-1}\dbar{p+2,p+2,n,n+2}\qquad&\rightarrow\qquad\text{spin 0,1},\\
	u^{p-1}\dbar{p+3,p+2,n,n+3}\qquad&\rightarrow\qquad\text{spin 0,1,2}.
\end{split}
\end{align}

\end{document}